# Stable and Semi-stable Sampling Approaches for Continuously Used Samples

Nikita Astrakhantsev[†]
Microsoft
Bellevue, USA
niastrak@microsoft.com

Deepak Chittajallu
Microsoft
Bellevue, USA
dechitta@microsoft.com

Nabeel Kaushal
Microsoft
Bellevue, USA
nabeelk@microsoft.com

Vladislav Mokeev
Microsoft
Bellevue, USA
vlamok@microsoft.com

## ABSTRACT

Information retrieval systems are usually measured by labeling the relevance of results corresponding to a sample of user queries. In practical search engines, such measurement needs to be performed continuously, such as daily or weekly. This creates a trade-off between (a) representativeness of query sample to current query traffic of the product; (b) labeling cost – if we keep the same query sample, results would be similar allowing us to reuse their labels; and (c) overfitting caused by continuous usage of same query sample. In this paper we explicitly formulate this tradeoff, propose two new variants – Stable and Semi-stable – to simple and weighted random sampling and show that they outperform existing approaches for the continuous usage settings, including monitoring/debugging search engine or comparing ranker candidates.

## 1 Introduction

Sampling – the process of selecting a small subset from a larger set (population) in order to infer some knowledge about this population – is usually considered to be a one-time action: once a dataset is sampled, it is assumed fixed. However, depending on the task nature, the sampled dataset may be used continuously: for example, if we want to measure quality of the search engine daily (or weekly, or with any other period – hereinafter we assume daily measurement), then we usually keep the same queries and only update results such as web documents that our system returns at each particular day. Since most of the returned results will be the same for a few days (and usually even for weeks and months), we can reuse their labels and reduce labeling cost. For example, if we reuse labels for a one thousand query sample in Bing image search for a month, we could save ~95% judgments over the month compared to rejudging each day. At the same time this creates other problems: (1) *staleness* – after some time, like 2-3 months, query set would not be representative for the queries issued in these last 2-3 months – in other words, the underlying population of all user issued queries changes, but sample remains the same; (2) *overfitting* - if this query set is used to decide if a candidate for system improvement should be shipped or not, which is usually the case, then after a few iterations the system would overfit to the query set, so that the measurement results would be an overestimation. Alternative solution – to change query set completely each day or each month – would solve these problems, but would require too many new labels, which are usually costly to obtain.

If the sampling needs to be not uniform, i.e. Simple Random Sampling (SRS), but weighted (Weighted Random Sampling, WRS), then the task becomes even more complicated. In this paper we propose a method to adjust query set each day (or each week/month/etc.), so that the sample (SRS or WRS) would be representative to the most recent query distribution and at the same time would share as much queries as possible with previous samples to reduce labeling cost; also, this method would allow gradually updating the query set, i.e., provide full control on the trade-off between labeling efforts and refreshing timeline, or proneness to overfitting. To the best of our knowledge, there is no other sampling algorithm for continuously used datasets that would have such properties.

This paper is organized as follows: Section 2 briefly reviews existing works on sampling, with focus on Efraimidis' and Spirakis' method for weighted random sampling [1]. In Section 3 we describe our Stable and Semi-stable approaches for simple and weighted random sampling. Section 4 provides experimental results. In Section 5 we compare existing approaches to sampling with Stable and Semi-stable ones.

## 2 Related work

Given the ubiquitous role of sampling in statistical analysis, it is not surprising that there is a plethora of literature on sampling, from Cohran's "Sampling techniques" [2], originally published in 1953, to a more recent works by Lohr [3], Tile [4], Fuller [5], etc.

With the increase of data (i.e., population) size, more works started to focus on effective methods to do sampling. For example, Meng [6] suggested a scalable SRS, which can be applied in a distributed setting with reduced storage and support of load balancing. Sanders et al. [7] optimize sampling methods for cache efficiency and number of communications between processors; their method supports modifications for online sampling, load balancing and vectorization needed for running on GPUs.

One important type of sampling is weighted random sampling (WRS). Here, each sampling unit comes with an associated weight and a weighted random sample should accurately represent the weight distribution of a population – so, if unit X has a weight of 2 and Y has a weight of 1, then X is twice as likely to be selected in a weighted random sample as Y.

Efraimidis and Spirakis [1] suggested an elegant algorithm to do weighted random sampling without replacement; given that our method is based on this work (and most useful for WRS), we will discuss this in more details and provide illustration in Table 3. Efraimidis' and Spirakis' method generates a uniform random number for each sampling unit (such as user query in search engine), then raises it to the power of inverse of weight (number of Impressions, in our case) and uses this final number as an order key, i.e., the user of sampling takes needed number of sampling units with biggest values of order keys.

This method is simple and parallelizable; it also supports arbitrary sample size: we can just take a prefix of bigger or smaller size. Worth noting that simply combining 2 valid weighted samples into 1 doesn't create a valid weighted sample; at the same time, preparing a bigger sample than needed and (uniform) subsampling from this would also create sample biased towards uniform sampling, cf. "Comparison results are obtained for the inclusion probabilities in some unequal probability sampling plans without replacement. For either successive sampling or Hajek's rejective sampling, the larger the sample size, the more uniform the inclusion probabilities in the sense of majorization." [8]

Chao [9] suggested an algorithm for Weighted random sampling with replacement, which is also belong to reservoir sampling family, as the method of Efraimidis and Spirakis.

As with SRS, there are multiple works focused on modifying WRS for different computational settings, e.g. Hübschle-Schneider and Sanders [10] optimized WRS (with and without replacement) for parallel settings. At the same time, to the best of our knowledge, sampling algorithms were not previously modified for settings with continuously changing populations.

Other related directions of research include choosing subset of query-document pairs to send for judgments [11, 12, 13]; combining samples from different distributions [14, 15]; and counterfactual evaluation, i.e., evaluating policies on offline logged data as opposed to (or before) online A/B tests, in particular the work from [16] that proposes estimators that can combine logs from multiple previous A/B tests. Compared to these approaches, the one proposed in this paper is agnostic to the documents, i.e., considers only queries and their weights, is simpler and can be applied to the datasets of any size.

## 3 Stable and Semi-stable approaches

As outlined in the introduction, we consider here the following two requirements to the sampling approach:

*Ability to update set regularly* (hereinafter let's assume monthly update) based on the most recent query logs, so that the overlap between previous set and new set is maximal. Let's call a sampling approach with this ability ***Stable***.

*Ability to update set monthly with overlap of controllable size*, so that we can explicitly control the trade-off between labels reusage and proneness to overfitting. Let's call a sampling approach with this ability ***Semi-stable***.

Below we describe Stable and Semi-stable sampling approaches for Simple random sampling (mostly, for illustration purposes) and Weighted random sampling.

### 3.1 Simple Random Sampling

If we want uniform or simple random sampling (SRS), i.e. treat each query with the same weight, then the method to sample N queries can be straightforward: assign uniform random numbers to each query, then rank by this number and take top N queries, see Table 1 (note that all queries there are synthetic and provided just for illustration purposes).

*3.1.1 Stable SRS.* Assume that we have 2 distributions here: with queries from May and from June. Then some queries would occur in both months; some queries – only in May; some – only in June, see Table 2. Stable SRS then would generate random number for each query in June as follows:

1) If a query occurred only in May – remove from the sample;
2) If a query occurred both in May and in June – take random number from May;
3) If a query occurred only in June – generate new random number.

Then just take top N queries based on these new random numbers (a lot of them would be still from May, though, assuming no drastic change in query distribution from May to June).

The final sample is a valid simple random sample, because each query has a random number generated independently from a uniform distribution.

As we can see, this new sample would contain some of the queries that occurred only in June and won't contain queries that occurred only in May; the queries that occurred both in May and June that were selected in May are likely to be selected in June as well, since the largest random numbers in May population are likely to be among the largest random numbers in June population.

*3.1.2 Semi-stable SRS.* If we want to be able to change some part of the sample – say, 10% each month – to avoid overfitting, then we can randomly split the whole population into 10 parts, choose one part and regenerate random numbers for it; at the next month, for the 3rd sample, we can regenerate random numbers for another part, see Table 3.

In this illustration, query "catastro" that had big random number previously (0.988), got small regenerated number (0.45) and thus was excluded from the set of top 7 queries; query "Another query" is also from the part where we regenerated random numbers and it got a high number, which led to its inclusion into the sample. Other queries in the sample are from the parts where we didn't regenerate numbers.

Note that if we have a set of random numbers, then randomly choose a subset and regenerate random numbers for this subset – after this process, the whole set would still have uniformly distributed random numbers. Therefore, Semi-stable SRS is again a valid simple random sample.

**Table 1. Illustration of SRS**

| Query | aasdfd | Ok asd | what is bing | Why use bing not gogle | catastro | Who killed jdsdf | need 1 more query |
|---|---|---|---|---|---|---|---|
| random number, u | 0.995 | 0.992 | 0.99 | 0.989 | 0.988 | 0.987 | 0.985 |
| Sampling order | 1 | 2 | 3 | 4 | 5 | 6 | 7 |

**Table 2. Illustration of Stable SRS**

| Query | aasdfd | Ok asd | what is bing | Why use bing not gogle | catastro | new query in June | Who killed jdsdf |
|---|---|---|---|---|---|---|---|
| random number, u (generated in May) | 0.995 | 0.992 | 0.99 | 0.989 | 0.988 | - | 0.987 |
| Sampling order - May | 1 | 2 | 3 | 4 | 5 | | 6 |
| random number, u (generated in June) | - | - | 0.99 | 0.989 | 0.988 | 0.9875 | 0.987 |
| Sampling order - June | | | 1 | 2 | 3 | 4 | 5 |

**Table 3. Illustration of Semi-stable SRS**

| Query | aasdfd | Ok asd | what is bing | Another query | Why use bing not gogle | catastro | new query in June |
|---|---|---|---|---|---|---|---|
| random number, u (generated in May) | 0.995 | 0.992 | 0.99 | 0.34 | 0.989 | 0.988 | - |
| Sampling order - May | 1 | 2 | 3 | 75913 | 4 | 5 | |
| random number, u (generated in June) | - | - | 0.99 | 0.9893 | 0.989 | 0.45 | 0.9875 |
| Sampling order - June | | | 1 | 2 | 3 | 6924913 | 4 |

**Table 4. Illustration of Efraimidis and Spirakis' algorithm for WRS**

| Query | cat pics | what is bing | dogs | images | aasdfd | mars | general | need 1 more query |
|---|---|---|---|---|---|---|---|---|
| Impressions, w | 123124 | 12 | 233242 | 456423 | 2 | 34 | 3425 | 1 |
| random number, u | 0.7 | 0.99 | 0.6 | 0.4 | 0.995 | 0.91 | 0.8 | 0.99 |
| Order key, $k = u^{1/w}$ | 0.992 | 0.989 | 0.976 | 0.974 | 0.969 | 0.966 | 0.961 | 0.960 |
| Sampling order | 1 | 2 | 3 | 4 | 5 | 6 | 7 | 8 |

**Table 5. Illustration of Stable WRS**

| Query | cat pics | what is bing | dogs | images | aasdfd | mars | generalization | need 1 more query | 1 more query from June |
|---|---|---|---|---|---|---|---|---|---|
| random number, u | 0.7 | 0.99 | 0.6 | 0.4 | 0.995 | 0.91 | 0.8 | 0.99 | 0.99 |
| Impressions in May, w | 123124 | 12 | 233242 | 456423 | 2 | 34 | 3425 | 1 | 0 |
| Order key for May, k = $u^{1/w}$ | 0.992 | 0.989 | 0.976 | 0.974 | 0.969 | 0.966 | 0.961 | 0.960 | - |
| Impressions in June, w | 124565 | 10 | 334242 | 455210 | 1 | 47 | 3400 | 0 | 1 |
| Order key for June, k = $u^{1/w}$ | 0.992 | 0.979 | 0.986 | 0.975 | 0.869 | 0.986 | 0.958 | - | 0.960 |
| Sampling order for June | 1 | 4 | 2 | 5 | 8 | 3 | 6 | - | 7 |

**Table 6. Illustration of Semi-stable WRS**

| Query | cat pics | what is bing | dogs | images | aasdfd | mars | generalization | need 1 more query | 1 more query from June |
|---|---|---|---|---|---|---|---|---|---|
| random number, $r_1$ (May) | 0.7 | 0.99 | 0.6 | 0.4 | 0.995 | 0.91 | 0.8 | 0.99 | - |
| random number, $r_2$ (June) | **0.8** | 0.99 | 0.6 | **0.12** | 0.995 | **0.68** | 0.3 | 0.45 | **0.99** |
| Impressions in May, w | 123124 | 12 | 233242 | 456423 | 2 | 34 | 3425 | 1 | 0 |
| Order key for May, k = $r_1^{1/w}$ | 0.992 | 0.989 | 0.976 | 0.974 | 0.969 | 0.966 | 0.961 | 0.960 | - |
| Sampling order for May | 1 | 2 | 3 | 4 | 5 | 6 | 7 | 8 | - |
| Impressions in June, w | 124565 | 10 | 334242 | 455210 | 1 | 47 | 3400 | 0 | 1 |
| Stable Order key for June, k = $r_1^{1/w}$ | 0.992 | 0.979 | 0.986 | 0.975 | 0.869 | 0.986 | 0.958 | - | 0.960 |
| Semi-stable Order key for June, k = $r_2^{1/w}$ | **0.997** | 0.979 | 0.986 | **0.972** | 0.869 | **0.654** | 0.958 | - | 0.960 |
| Semi-stable Sampling order for June | 1 | 3 | 2 | **4** | 47235 | **864345** | 6 | - | 5 |

## 3.2 Weighted Random Sampling

Let's start from illustration of ordinary Weighted Random sampling: in Table 4, for each query there (1st row), we have number of impressions *w* (2nd row) and randomly generated number *u* (3rd row). The last row contains a final order key computed from Impressions and random number by simple formula. If we need to take sample of size 5, we'll take queries "cat pics", "what is bing", "dogs", "images" and "aasdfd", because they have biggest values of order key.

*3.2.1 Stable WRS.* The underlying idea of Stable WRS is a nice property of Efraimidis and Spirakis' algorithm: If we change weights but keep the same original random numbers (2nd row in Table 4), then we can get a valid sampling order for these new weights. The reason is that the original set of random numbers is not worse than any other set of random numbers, e.g. specifically generated for the 2nd sample. Thus, we can apply the same strategy to random numbers as we do for Stable SRS and use order key computed by Efraimidis and Spirakis' formula instead of ordering by random numbers, see Table 5. In this example, we'll have slightly different order of queries for June impressions compared to May impressions - e.g. "dogs" will be second, while "what is bing" will be third, because of difference in impressions; some low-popular queries will disappear and be replaced by new low-popular queries; some queries that occurred only in May will also disappear; and so on.

Note that this works not only for the query sets from different points in time – like, from May and June, or from last year and current year – but also for query sets accumulated from different sources, e.g. from different regions or different parts of search engine (web search results page, or SERP, vs image search results page, or IRP) or even different search engines (Bing vs Google vs Yandex). As long as queries overlap between 2 sources and have similar weights ordering numbers (e.g. in SERP queries "google" and "cats" may have 10M and 1M impressions, while in IRP the same queries may have 1M and 100k impressions, which is different, but the ordering number may be quite similar – around 1st and 100th) for a significant number of queries, we can have 2 valid samples with big overlap, thus reducing number of new judgements. Note that judgments required for web search vs image search are likely to be different in general case, but some type of labels like query intent can be shared between web and image searches.

In a sense, this "reweighting" operation provides us with a possibility to make "views" on sample, so that each view is a valid sample for the corresponding population (set of weights), but at the same time overlaps with something else – as noted above, this makes sense to do as long as 2 populations, i.e. orderings of queries by weights, are similar. Of course, if orderings by weights are not similar, then samples would have small overlap, but would still be valid.

*3.2.2 Semi-Stable WRS.* As with SRS, we can change random numbers for a subset of original random numbers, see Table 6. According to Stable WRS, we add random number for "1 more query from June" and we would use Order key from the penultimate row. According to Semi-stable WRS, we randomly change random numbers for queries "cat pics", "images", "mars" (see Order key in the last row). New random number doesn't change position of "cat pics" and "images", because they have too big impression counts, but query "mars" disappears from the June sample.

## 3.3 Hashing trick

As shown above, both weighted random sampling and simple random sampling approaches can be Stable or Semi-stable depending on how we generate and update random numbers. We can store these random numbers for each query in the population, but this may require a significant storage volume, plus for semi-stable approach we must store these random number sets for the whole population at each sampling date, which may quickly become unmanageable, especially if we want to share this between multiple teams.

Instead, we can rely on the uniformity of a good hashing function, e.g. apply md5 to the query string itself – concatenated with some constant text string – let's call it seed, as in pseudo-random number generators - to be able to generate different random numbers.

For Stable sampling, Hashing trick is straightforward – we just store hashing seed. For Semi-stable sampling, we propose the following algorithm. It is based on the idea to have 2 different hashing functions: one would be used to generate random numbers as in Stable approach, while the other would decide for each query if new hashing function should be used, i.e., if this query should keep random number from previous sample or should have a new one.

**Pseudocode for Hashing trick.**

**Input:**
*Operations:*
  1) SampleHash(Seed, Query) // Hashing function that generates random number based on some Seed and Query, e.g. md5 from concatenation of Seed and Query
  2) NewSeed() // Function that generates new Seed each time it is called, e.g. text string Month+Year+Constant if we generate new seeds not more often than each month; or just an iterator on prepopulated list of Seeds.
  3) RefreshHash(Seed, Query) // Hashing function that generates random number based on some Seed and Query, but independent from SampleHash, e.g. md5 from concatenation of Seed, Query and some constant string

*Constants:*
  1) S1, S2   // 2 start seeds, e.g. some unique identifier of the current sample and the first call to NewSeed()
  2) DesiredRefresh  // float number specifying desired change in random numbers, e.g. if we want to update the whole sample completely after 12 times, then we should set it to 1/12=0.0833

**Output**: Random numbers R for each Query

Initialize:
  1. Seed = (S1, S2)
  2. Refresh = 0.0

Each Rolling Period:
  1. Refresh = Refresh + DesiredRefresh
  2. IF Refresh > 1
       a. (S1, S2) = (S2, NewSeed())  // Rolling Seed
       b. Refresh = Refresh % 1
  3. FOREACH Query:
     R = (RefreshHash(S1, Query) > Refresh) ? SampleHash(S1, Query) : SampleHash(S2, Query)

## 4 Experiments

We performed simulation experiments on Bing query logs collected from 2019 to 2020 on main page (Search Engine Results Page, SERP) and image search page (Image Results Page, IRP).

For all samples, we used the following parameters:

| Parameter | Value |
|---|---|
| *Sample size* | 1000 |
| *Sampling time interval* | 12 months |
| *Rolling frequency* | 1 month |
| *Refresh (for semi-stable approach)* | 10% |

### 4.1 Illustration of sample validity

Here we illustrate that Stable and Semi-stable approaches provide valid samples by plotting Cumulative distribution function (CDF) for population, where we consider impression volume (i.e. sum of impressions for all queries), and for samples produced by each approach, where we consider count volume (i.e. count of queries, ignoring their impressions) – intuition here is that if the original population have ~55% of all impression volume accumulated by queries with impression lesser than 10, then ~55% of our sample should contain queries with impressions lesser than 10.

**Table 7. CDF plots for Stable and Semi-stable WRS**

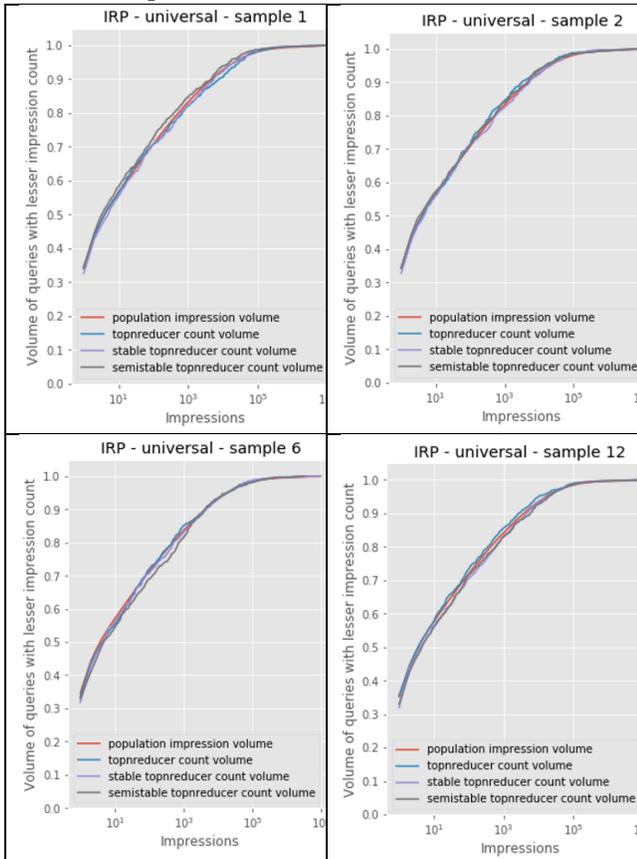

As we can see, samples correspond to population in the 1$^{st}$, 2$^{nd}$, 6$^{th}$ and last months.

## 4.2 Estimating overlap

Here we measure actual overlap between the first sample and each consecutive sample for different sampling approaches (for original WRS proposed by Efraimidis and Spirakis [1], overlap between any two independent samples is less than 1% in all cases, therefore it is not plotted here).

As we can see from Figures 1 and 2, overlap depends on desired refresh coming from Semi-stable sampling and on natural churn occurring due to changes in queries (old queries unique to dropped months, new queries unique to current month) and weights – see dependence in Table 8. Note that Natural retention/churn depends on the weights distribution; for IRP it was empirically found to be around 0.93, which is close to 11/12=0.9167 – proportion of the overlapping months in 1-year sample. Intuitively, the proximity to 11/12 reflects the tail-heavy nature of IRP traffic - it nearly acts as a population of seen-only-once queries (for which monthly churn would be 1/12). The actual number being higher reflects the fact that queries do in fact repeat. Less taillish distributions - such as SERP - are expected to have even higher natural retention, see Figure 3.

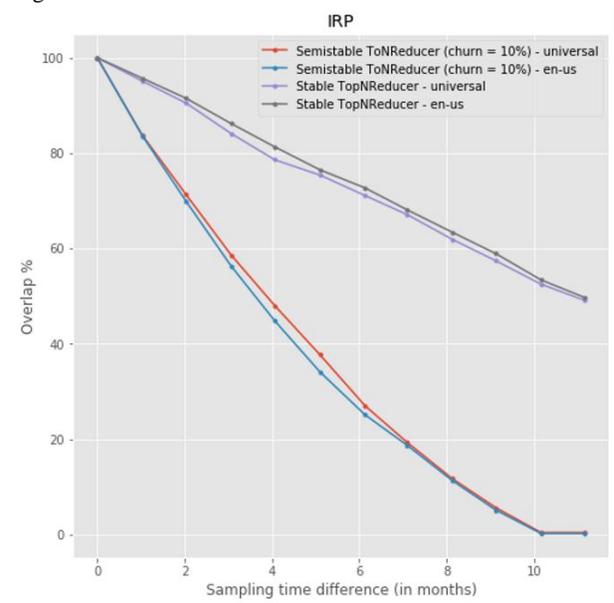

**Figure 1. Desired refresh 10% on Image Results Page**

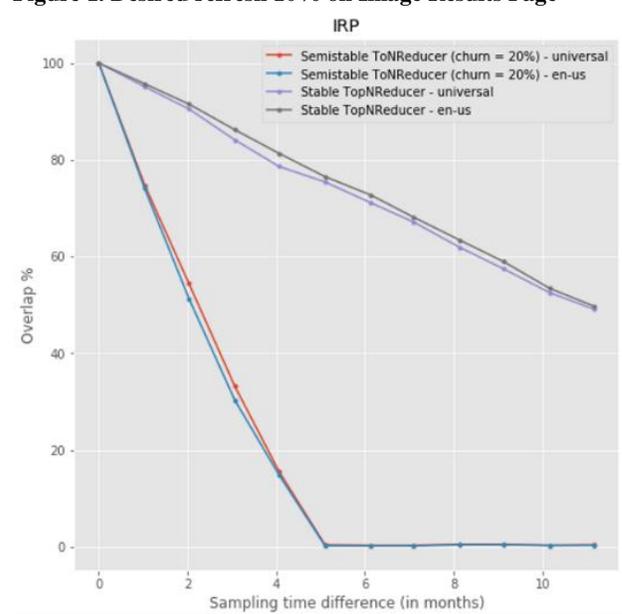

**Figure 2. Desired refresh 20% on Image Results Page**

Stable and Semi-stable Sampling Approaches for Continuously Used Samples

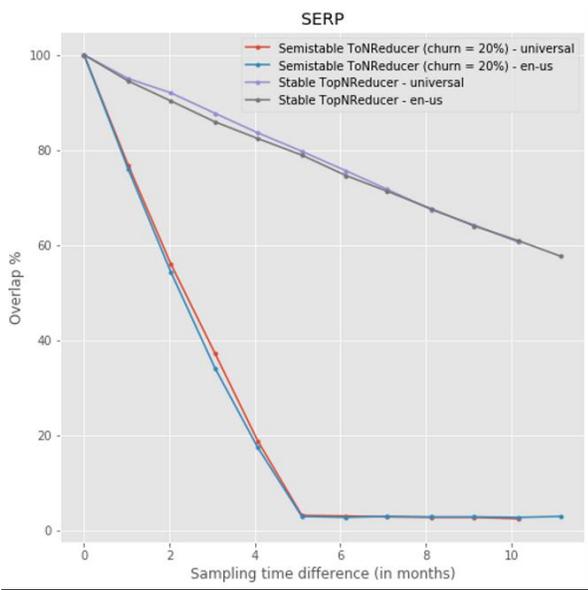

Figure 3. Desired refresh 20% on Search Engine Results Page

From Table 8 we can also see that mean overlap for IRP is smaller than for SERP.

Table 8. Averages of consecutive overlaps

| Source | Method | Mean consecutive sample overlap |
|---|---|---|
| IRP | WRS | 0.24% |
| IRP | Stable WRS | 94.1% |
| IRP | Semi-stable WRS (refresh 10%) | 85.0% |
| IRP | Semi-stable WRS (refresh 20%) | 75.0% |
| SERP | WRS | 3.1% |
| SERP | Stable WRS | 95.1% |
| SERP | Semi-stable WRS (refresh 20%) | 77.0% |

See Table 9 for statistics on relations between Semi-stable Refresh, Natural churn (proportion of queries to be changed from month to month for an ordinary sampling).

Note that month-to-month delta in Final overlap is decreasing – from 0.15 to 0.13 to 0.04 at the end – but judgment load remains essentially constant after the first month; it equals *NaturalChurnRate + Refresh - (NaturalChurnRate \* Refresh)*, where the last component is almost zero.

In practice, it is also useful to know actual judgment load, i.e. the number of queries to be judged each month, assuming reusage of any previous judgment. Final Overlap just compares first query sample to current query sample, but queries can rechurn - causing judgment load but not affecting overlap.

Table 9. Relations between Semi-stable Refresh, Natural Retention (NatRet), Natural churn and Overlap

| Month Delta | Natural Retention | Natural churn | Refresh (in Semi-stable) | Refresh churn | Natural + Refresh churn | Final Overlap |
|---|---|---|---|---|---|---|
| d | $(11/12)^d$ | 1 - NatRet | d/12 | Refresh * NatRet | | |
| 0 | 1.00 | 0.00 | 0.00 | 0 | 0.00 | 1 |
| 1 | 0.93 | 0.07 | 0.08 | 0.0775 | 0.15 | 0.85 |
| 2 | 0.86 | 0.14 | 0.17 | 0.1442 | 0.28 | 0.72 |
| 3 | 0.80 | 0.20 | 0.25 | 0.2011 | 0.40 | 0.6 |
| 4 | 0.75 | 0.25 | 0.33 | 0.2494 | 0.50 | 0.5 |
| 5 | 0.70 | 0.30 | 0.42 | 0.2899 | 0.59 | 0.41 |
| 6 | 0.65 | 0.35 | 0.50 | 0.3235 | 0.68 | 0.32 |
| 7 | 0.60 | 0.40 | 0.58 | 0.3510 | 0.75 | 0.25 |
| 8 | 0.56 | 0.44 | 0.67 | 0.3731 | 0.81 | 0.19 |
| 9 | 0.52 | 0.48 | 0.75 | 0.3903 | 0.87 | 0.13 |
| 10 | 0.48 | 0.52 | 0.83 | 0.4033 | 0.92 | 0.08 |
| 11 | 0.45 | 0.55 | 0.92 | 0.4126 | 0.96 | 0.04 |
| 12 | 0.42 | 0.58 | 1.00 | 0.4186 | 1.00 | 0 |

See Figures 4 and 5 that estimate number of queries to be judged at each month, assuming reusage of previous judgments.

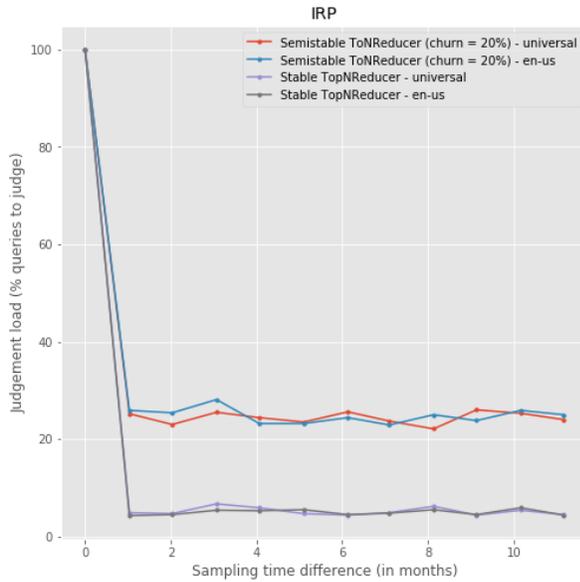

**Figure 4. Judgment load for Image Results Page**

As we can see from these figures, judgment load is almost constant after the first month, which is a desirable property in practice.

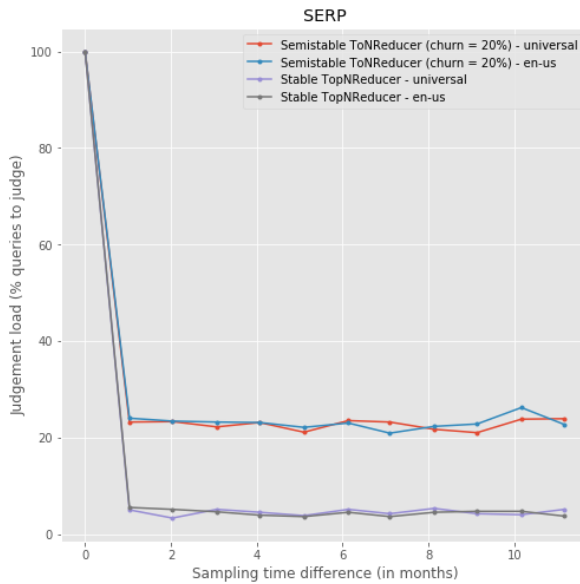

**Figure 5. Judgment load for Search Engine Results Page**

## 5   Comparison of sampling approaches

As noted in the Introduction above, there may be multiple sampling approaches for continuously changing populations. To the best of our knowledge, they are the following (see also Table 10 below for a summary).

**Keep same set forever** – the simplest solution is to just create one sample (using original WRS) and use it for the whole project lifetime. This obviously creates a valid (first and only) sample for the population at the start of the project and labels can be reused as much as possible, but with time this sample would go out of sync with the population if it changes fast enough and overfitting on this set is also inevitable with time, even for static populations. Therefore, this extreme approach is good only for short projects, when problems won't have time to accumulate.

**Change sample as soon as possible** – another extreme approach is to change the whole sample with each usage. This completely solves freshness and overfitting problems, but it requires too many judgments, which in practice would lead to too small sample sizes due to budget constraints, which, in turn, would lead to statistically wider noise level in search quality metrics like NDCG preventing us from being able to detecting smaller improvements to the search engine.

**Keep sample for multiple periods, then change** – this approach is a combination, or generalization, of the previous two: instead of changing each day or keeping forever, we keep the sample for some number of usages (e.g. for a month or a year) and then resample. This provides a valid sample with reasonable reuse of judgments, thus it is probably the most popular solution, but it has a few problems as well: (a) the sample becomes stale approaching the end of each month/year (period of resample); (b) there is usually enough time to overfit, so that results of the system on the resampled set becomes worse than on the previous day – in the worst case, results may be even the same as they were at the start of the previous sample; (c) judgment systems usually prefer stable flow of small tasks to the rare spikes of huge tasks, therefore resampling causes lower efficiency of labeling and delays in getting labeled data.

**Stable sampling** – as described above, this approach prioritizes freshness and judgments reuse; at the same time, given that most of the queries remain the same, the overfitting resistance of this approach is low - just slightly better than that of the 1st and the 3rd approach due to a side-effect of freshness: some queries would change due to changes in underlying distribution.

**Semi-stable sampling** – as described above, this approach is designed to be valid and fresh and to provide full control on the trade-off between overfitting resistance and judgments reuse. Note also that here we can sync desired refresh, i.e. the main hyperparameter of the method controlling which part of the sample would be refreshed each time, with the lifespan of the label – e.g. if we decide that after a year the same pair of query and result needs to be rejudged, we can set desired refresh so that the sample would be fully refreshed in a year; we can also adapt to judgment budget changes quickly: if for month or two we have smaller judgment budget, we can temporarily reduce desired refresh until situation normalizes and then temporarily increase refresh to gradually catch up.

**Replace random subset of sample** – this approach tries to gradually update the sample by randomly choosing a subset and replacing it by a new sample of the same size as the subset. There may be multiple variations depending on how to determine the subset to replace at each time: e.g. it is possible to split the set into halves each month independently and replace one half with a new sample; alternatively, split can be made just once and then older part can be replaced each month. For Simple Random Sampling, this is a valid approach and similar to Semi-stable sampling, without updates to a new population each month; however, for Weighted Random Sampling, this does not produce a valid sample. In particular, such methods require a strategy on what to do with duplicates – given that the next sample is independent on the previous ones, it is probable that some queries (usually, head ones, i.e. occurring most often) would occur in both previous and next samples; if we just ignore the problem, then our sample would be biased toward tail queries, i.e. occurring less often; if we choose to replace query by another query with similar weight, we would also bias the sample (why this particular query, how to choose if there are multiple queries with the same weight, etc.).

To sum up, this approach is more complicated, provides weighted sample with worse freshness than Semi-stable approach and the validity of such sample is not proven.

Table 10 compares approaches by the following properties:

1. **Weighted sample validity**: binary property showing if it is proved that the sampling approach produces valid WRS sample at each time; in particular, it shows if any statistic like mean NDCG computed on the sample can be generalized to the original population.
2. **Freshness**: for continuously changing populations like user queries, how fresh the samples are – for example, if we keep the same sample for a year, then after a few months the sample isn't fresh and thus may be not representative for the latest population.
3. **Resistance to overfitting**: a measure of sample static, or, from practical point of view, how prone to overfitting the system based on such sampling would be.
4. **Judgment bandwidth savings**: a measure of number of new labels required.

## 6 Conclusion

In this paper, we highlighted the key practical requirements of sampling approaches for continuously used datasets like query sets used to measure search engines daily. We show that current sampling approaches have drawbacks and propose a new Semi-stable approach for both Simple and Weighted Random Sampling that is provably valid and provides full control on the trade-off between judgments reuse and proneness to overfitting. We experimentally evaluate this approach on real data (query logs of Microsoft Bing) and show that it creates valid samples with specified overlap between consecutive samples.

One of the limitations of the presented Stable and Semi-stable approaches is inability to use labeled data for training, because any query can be reused in future refreshes. How to account for necessity of training data is one of the possible directions for future work.

## REFERENCES

[1] Efraimidis, Pavlos, and Paul Spirakis. "Weighted Random Sampling: 2005; Efraimidis, Spirakis." Encyclopedia of Algorithms (2008): 1024-1027 (pdf)
[2] Cochran, William G. Sampling techniques. John Wiley & Sons, 2007.
[3] Lohr, Sharon L. Sampling: design and analysis. Nelson Education (2009).
[4] Tille, Yves. Sampling Algorithms - Springer. Springer Series in Statistics. doi:10.1007/0-387-34240-0. ISBN 978-0-387-30814-2.
[5] Fuller, Wayne A. Sampling statistics. Vol. 560. John Wiley & Sons (2011).
[6] Meng, Xiangrui. "Scalable simple random sampling and stratified sampling." In International Conference on Machine Learning, pp. 531-539 (2013).
[7] Sanders, Peter, Sebastian Lamm, Lorenz Hübschle-Schneider, Emanuel Schrade, and Carsten Dachsbacher. "Efficient Parallel Random Sampling—Vectorized, Cache-Efficient, and Online." ACM Transactions on Mathematical Software (TOMS) 44, no. 3 (2018): 1-14.
[8] Yu, Yaming. "On the inclusion probabilities in some unequal probability sampling plans without replacement." Bernoulli 18.1 (2012): 279-289. https://arxiv.org/pdf/1005.4107.pdf
[9] Chao, Min-Te. "A general purpose unequal probability sampling plan." Biometrika 69, no. 3 (1982): 653-656.
[10] Hübschle-Schneider, Lorenz, and Peter Sanders. "Parallel weighted random sampling." arXiv preprint arXiv:1903.00227 (2019).
[11] Allan, James, Ben Carterette, Javed A. Aslam, Virgil Pavlu, Blagovest Dachev, and Evangelos Kanoulas. "Overview of the TREC 2007 million query track." In Proceedings of TREC (2007).
[12] Carterette, Ben, Virgil Pavlu, Evangelos Kanoulas, Javed A. Aslam, and James Allan. "If I had a million queries." In European conference on information retrieval, pp. 288-300. Springer, Berlin, Heidelberg (2009).
[13] Yilmaz, Emine, Evangelos Kanoulas, and Javed A. Aslam. "A simple and efficient sampling method for estimating AP and NDCG." In Proceedings of the 31st annual international ACM SIGIR conference on Research and development in information retrieval, pp. 603-610 (2008).
[14] Elvira, Victor, Luca Martino, David Luengo, Monica F. Bugallo. Efficient multiple importance sampling estimators. IEEE Signal Processing Letters 22, 10, pp.1757–1761 (2015).
[15] Elvira, Victor, Luca Martino, David Luengo, and Monica F. Bugallo. Generalized multiple importance sampling. arXiv preprint arXiv:1511.03095 (2015).
[16] Agarwal, Aman, Soumya Basu, Tobias Schnabel, and Thorsten Joachims. "Effective evaluation using logged bandit feedback from multiple loggers." In Proceedings of the 23rd ACM SIGKDD International Conference on Knowledge Discovery and Data Mining, pp. 687-696 (2017).

**Table 10. Comparison of Sampling approaches**

| # | Sampling approach | Weighted sample validity | Freshness | Resistance to overfitting | Judgment bandwidth savings | Notes |
|---|---|---|---|---|---|---|
| 1 | Keep same sample forever | + | Lowest | Lowest | Highest | Fine for short projects |
| 2 | Change sample ASAP | + | Highest | Highest | Lowest | |
| 3 | Keep sample for a year, then change completely | + | Low | Low | Med | Huge labeling spike when change set: not friendly for label collection systems |
| 4 | Stable sampling | + | High | Low-Med | High | Best for monitoring/debugging |
| 5 | Semi-stable sampling | + | High | (Controllable) | (Controllable) | DesiredRefresh can consider judgements lifespan |
| 6 | Replace random subset of sample | - | Med | (Controllable) | (Controllable) | |